\documentclass[journal]{newIEEEtrans}

\usepackage{amsmath, amssymb, enumerate}
\usepackage{fancybox}
\usepackage{amsfonts}
\usepackage{algorithm}
\usepackage{algorithmicx}
\usepackage{algpseudocode}
\usepackage[usenames]{color}
\usepackage{listings}
\usepackage{graphicx,dblfloatfix}
\usepackage{times}
\usepackage{psfrag}
\usepackage{subfigure}
\usepackage{caption}
\usepackage{multirow}
\usepackage{epsfig}
\usepackage{url}
\usepackage{color}
\def\fixme#1{\typeout{FIXED in page \thepage : {#1}}
\bgroup \color{red}{[FIXME: {#1}]} \egroup}
\usepackage{rotating,tabularx}

\begin{document}

\title{A Proposed System for Covert Communication to Distant and Broad Geographical Areas}
\author{
Joshua Davis\\
cwstats@covert.codes\\
}

\maketitle
\thispagestyle{empty}

\begin{abstract}
A covert communication system is developed that modulates Morse code characteristics and that delivers its message economically and to geographically remote areas using radio and EchoLink.  Our system allows a covert message to be sent to a receiving individual by hiding it in an existing {\em carrier} Morse code message.  The carrier need not be sent directly to the receiving person, though the receiver must have access to the signal.  Illustratively, we propose that our system may be used as an alternative means of implementing {\em numbers stations}.
\end{abstract}

\section{Introduction}
A potential use for the system we will soon derive is a secure and economical alternative to numbers stations.  Numbers stations can be found throughout the high frequency (HF) radio band, as the propagation characteristics of this band allows them to operate over long distances and across broad areas.  The content of these stations consists of a seemingly random series of numbers or letters, in voice or in Morse code. It is speculated that these systems are operated by governments to allow them to communicate securely with covert human intelligence assets \cite{bbc}.  In this paper we make the assumption that the presumed purpose of numbers stations is correct, and suggest an improved, or at least alternative, way of doing the same thing.  If our presumptions are false, and numbers stations do not serve the role they are widely thought to serve, our system may still find utility in covert communication applications that would benefit from easy covert access to broad and remote geographical areas.  As with numbers stations, the receiver of our channel (such channels are generally simplex) may remain very anonymous.

Since our goal is to replace the functionality of numbers stations, our channel criteria (bandwidth, data-rate, level of security) should equal or exceed the level of these criteria found in numbers stations.  The carrier communication format we have selected for our covert channel is CW (Morse code).  This format is used frequently on HF bands for long range communication due to its resilience against noise, which is prevalent and somewhat unpredictable in long range HF communication, and its low bandwidth requirements when compared to Single-sideband (SSB) or Amplitude Modulation (AM).  Traditional numbers stations utilize SSB or Amplitude Modulation AM to communicate voice (synthesized or human), or CW to communicate Morse code.  Our channel uses CW and so occupies less than or approximately equal to the bandwidth used by numbers stations.  The data rate of numbers stations is presumably low, as the coded message is delivered one symbol (letter, number) at a time, at a rate suitable for manual human decoding.  Our system has a variable data-rate, and can emulate a human user, when provided information about a user's Morse code statistical behavior.  Numbers stations utilize simplex (one-way) communication, one-time pads, and long range radio to ensure that both the receiver and message are anonymous.  Our system may be utilized in this fashion, providing a level of security equal to numbers stations if this is the desired configuration.  Further, our system obfuscates to some degree the {\em existence} of the covert communication; something that numbers stations are not known to do.

\subsection{Morse Code}
Morse code is a simple and durable means of transmitting symbols (e.g. letters, numbers).  It is widely used by amateur radio operators on HF bands to communicate with one another over long distances.  In typical usage, a carrier of uniform amplitude and frequency is turned on and off to represent the Morse code symbols.  This is called Continuous Wave (CW) operation.  Morse code requires relatively little bandwidth at typical speeds \cite{W8XR}, and is noise tolerant.

Morse code consists of a standardized set of symbols, which are represented over the air by short {\em dots} and longer {\em dashes}.  For example, the letter {\em A} is represented by a dot, followed by a dash.  The duration of dots and dashes are proportional to the speed of the transmission, which is generally measured in Words per Minute (WPM).  A dot is defined as lasting one time unit.  Dashes last three time units.  The voids between elements (dots, dashes) in one symbol (e.g. letter) last one time unit.  The voids between symbols last three time units, and the voids between words are seven time units long.  Using Morse code, one may transmit any message they'd like via CW, assuming the receiver has access to the communication, generally over radio, and that the symbol set is shared between the transmitter and receiver.  In typical usage, operators use the established Morse code symbol set \cite{W1WC}.

\subsection{HF Communication}
HF radio communication is used by both amateur radio operators and governments to communicate over long distances, and across large geographic areas, as the properties of HF propagation are amicable to such usage.  HF is sometimes referred to as {\em shortwave}, especially in the context of broadcasting.  Shortwave broadcasting is sometimes done by government and news agencies (the BBC World Service being a prominent example).  One who tunes through the HF bands looking for interesting things to listen to, without the intention of participating themselves, is a {\em shortwave listener}.  HF radios are available at reasonable prices, and throughout the globe, so shortwave listening is a fairly popular hobby.  A numbers station recipient may easily obtain a shortwave receiver from a local electronics market, and the possession of one is not itself so unusual as to arouse suspicion.

Governments use HF for three primary purposes.  First, HF is used to link facilities and stations together.  As HF is generally not as fast or reliable as other forms of communication (e.g. the Internet, satellite communications), such systems are often used as backup or emergency systems.  Embassies often have large, very visible HF antennas on their roofs.  Second, governments use HF for propaganda.  HF broadcasting allows the transmitter to communicate with a distant and broad geographical area with a single antenna, which maybe located on friendly soil.  Those with shortwave radios may listen to uncensored news and propaganda anonymously and often in their native language.  Finally, HF is thought to be used to communicate with undercover human intelligence assets, via systems such as numbers stations.  An agent with a shortwave radio copies down a code at a predetermined time and day, and decodes the information using a pre-shared key (e.g. one-time pad).  When a one-time pad is used, the communication is thought to be entirely secure, as long as an adversary does not obtain access to the message recipient or the one-time-pad.

The quality of HF communication is dependent on several factors, some being the weather and the time of day.  To maintain relatively high levels of reliability over long distances, careful planning, powerful transmitters, and large antennas are sometimes necessary.  So, while HF serves a definite purpose, it is not always an optimal solution.  Our covert system may use HF, as in traditional numbers stations, but this need not be the case.  If HF resources are not accessible or optimal for the transmitting individual, a system such as EchoLink can be used to deliver the covert and encoded message to the recipient.

\subsection{EchoLink}
Echolink \cite{echolink} is a service that registered amateur radio operators can use for free.  EchoLink allows part (or all) of the communication path between amateur operators to occur over the Internet, using Voice over Internet Protocol (VoIP) protocols.  For example, an amateur operator, having registered with the service, may communicate with another {\em ham} across the globe, using their relatively short ranged Ultra High Frequency / Very High Frequency (UHF/VHF) hand-held radio, computer, or cell phone \cite{echolink_android}.  An EchoLink station will receive their radio communication, translate it to VoIP, and send it to the recipient EchoLink station, which will translate the communication back to Radio Frequency (RF) and send it over the air to the recipient.

There are thousands of EchoLink nodes currently deployed around the world, giving the user radio transmit capability to many geographic areas from the comfort of their living room (or military base, or secret facility).  The number of stations in a given area depends on the population density and standard of living in the area, as well as the amicability of the area's governing entity to amateur radio activity.  The EchoLink website \cite{echolink} allows anyone to locate nodes in a given area.  As of this writing, most nodes use VHF/UHF, though a few report to use HF, and so one may conceive that the advantages of HF propagation might be used in conjunction with EchoLink or a similar system \cite{IRLP} to allow the user to areas both far and broad.

\section{Our Channel}
With the above tools to some degree understood, we combine them to create a system that is part covert channel, part numbers station-style secured broadcast system.  That is, we devise a method to communicate over broad geographical areas, with the message obscured.  Even if the existence of the covert message is identified, it cannot be easily deciphered.  Our system is proof of concept: our channel has been developed and tested, but no significant mathematical analysis or derivation has been attempted, and legal restrictions limit our ability to test over amateur radio channels \cite{part97}.

Our channel as described here makes use of technologies such as EchoLink, VHF/UHF radio, and perhaps HF.  The concept is flexible enough that it may be used over existing operator infrastructure (e.g. the HF radio assets governments own and may already be using for traditional numbers stations), or other long-range radio networks such as the Internet Radio Linking Project (IRLP) \cite{IRLP}.  We use EchoLink here because it provides easy and economical access to many diverse geographical areas.  We would sometimes like to use HF over EchoLink, and would if such stations where available, though they seem to be rare or nonexistent at this time.

Our channel modulates CW, providing in effect {\em Morse code over Morse code}.  We do this by varying the statistical properties of the overt (carrier) Morse code communication.  As microcontroller assisted keying has become more common, the statistical properties of Morse code have become more uniform.  In particular, the standard deviation of carrier keying times have been lowered.  In our channel we are forced to make one more assumption: our carrier communication is presumably formed by a traditional electromechanical key, where the tone length (and associated statistics) are directly controlled by the time the key is depressed, which is directly controlled by the very imperfect (in terms of milliseconds) human operator.

Since our channel is Morse code modulated on Morse code, and we send an element (dot, dash) with each element of the carrier message, our channel speed is about the same as the carrier keying speed, measured in words per minute.  At this time, we have already seen the following desirable properties of our covert channel:

\begin{itemize}
\item Our channel is to some degree covert.  Observers see the carrier Morse code communication, whose contents are innocuous.  The carrier message can be formed naturally by communicating with an amateur radio operator, and by answering his questions or inquiring about his activities.  This other radio operator is unaware that our responses and inquiries carry covert data: to all receivers but our own decoding software, the carrier message is not unusual.

\item Our channel is sufficiently fast that it can be used for the purpose of sending coded characters ``to the field''.  That is, it operates at CW and its associated speeds, as do many numbers stations.

\item Our channel is noise resilient.  Morse code is used by amateur radio operators wanting to communicate over large distances in part because of this property.  Also, our channel relatively little bandwidth \cite{W8XR}.

\item Finally, our covert channel carrier is quite innocuous.  One may find many Morse code conversations over the amateur radio bands at any given time.  Adding our needle to this haystack won't raise any eyebrows, assuming we don't behave aberrantly on our carrier.
\end{itemize}

The channel transmitter and receiver share three things: knowledge of the time and location of transmission, a seed or use with a Pseudo-random Number Generator (PRNG), and the transmitter's base keying statistics.  The time and location of transmission is knowledge also required by recipients of traditional numbers stations; one may not decode what they cannot find.  Numbers stations also require the receiver to have the key, or one-time pad, available for decoding the message.  We suggest that our PRNG seed may double as a one-time pad, with arbitrary long pads derived by inputting the PRNG into a message digest algorithm.  It is conceivable that using the PRNG seed as a key may compromise the anonymity of the message, a consideration that must be weighed during the design of a particular channel implementation.

These previous two items, the time and location of transmission, and PRNG (one-time pad), are pre-shared both in our system, as well as in traditional numbers stations.  Our channel also requires the receiver to know some information about the transmitter's keying statistics.  This information can be pre-shared, or it can be derived during a pre-arranged {\em training period}, during which the transmitter communicates unmodulated messages to unwary amateur operators, and the receiver unobtrusively listens and derives statistics from the transmission.  The required statistics are the dot and dash time averages, and their standard deviations.  Note that in our example implementation, our deterministic transmitter (a Python script) emulates human statistical behavior by taking as input the averages and standard deviations of a human user's keying times.

With the transmitter and receiver aware of the element statistics of the pending conversation, and in possession of the seed, covert transmission may begin.  If EchoLink is used to economically carry the message, the transmitting individual selects an EchoLink node in the vicinity of the recipient.  The transmitter then calls {\em CQ}, and talks to whomever will answer.  The receiver records the transmitter's conversations, and when they have ended, decodes them.  For each element (dot, dash) of the overt transmit message, an element duration is taken from a normal distribution with the element average time as its mean.  As the transmitter and receiver PRNGs share the same seed, they both know what duration each element should have.  Deviation from the pre-shared durations convey the covert message.  The use of a shared PRNG seed and statistical distributions to aid in covert communication was discussed in my previous work on web read-time modulation \cite{Davis14_3}.

The symbols for the covert message are conveyed by shifting the pseudo-random symbol time up, perhaps by one standard deviation, to represent a covert dot, and down to represent a covert dash.  To represent the end of a letter, symbol time is left at its PRNG generated time.  As the averages and standard deviations are taken from real human CW messages, as a PRNG is used to generate the {\em particular} symbol times, and as the offsets are constrained to some maximum time (e.g. one standard deviation), the channel's symbol times appear innocuous to those who may happen to be quantifying the carrier message's statistics.  That is, observers do not have the PRNG seed, and they cannot determine whether the symbol times have gone up or down from the shared reference.

\section{Some Considerations}
The covert channel recipient must have decoding software.  If they are familiar with the implementation specifics, coding such a receiver is not prohibitively difficult.  It is speculated that poly-tone numbers stations exist \cite{slotmachine}.  Such stations have the potential to transmit information at a higher rate than stations based on voice or CW.  That they are not used exclusively is probably due to their increased bandwidth and lower noise resilience compared to voice and CW transmissions, and the necessity that the receiver posses specialized decoding software.

Like numbers stations, our channel operates in simplex (one-way) mode, and the recipient probably cannot be identified as the carrier traffic covers a broad area, which probably contains many individuals.  Security of the covert transmission is dependent on the strength of the encryption.  It is said that one-time pads provide {\em perfect} anonymity of a message, though this may be to some degree compromised if the PRNG doubles as the pad.  The existence of the covert communication is secure to the degree that realistic user behavior is emulated.  For manual electromechanical keying, the standard deviation of symbols should be high enough that we may modulate symbol time without arising suspicion, when a PRNG and normal distribution are used as previously mentioned.

EchoLink requires registration.  This suggests that an individual may be associated with the transmission.  Though this may seem dangerous, there are mitigating factors.  First, the covert message is difficult to identify, and so the transmitting individual should incur no suspicion.  Second, if the channel is used in the context of traditional numbers stations, which are presumably operated by governments, obtaining a false amateur radio license should not be a problem.  Finally, if it is determined that EchoLink is too risky for a particular transmission, other assets or networks may be used to transmit the covert message.  Other assets would also be used when there is not an EchoLink node in range of the recipient.

Multipath propagation may affect the duration of the signal at the receiver and undermine the accuracy of the channel.  To compensate, the code deviation time may be increased where multipath will cause problems.  One should examine to what extent increasing the code deviation undermines the statistical correctness, and thus covertness, of the channel.  Fortunately the statistics in question vary from individual to individual, so one may simply emulate a slower keyer when addressing this issue.  As standard deviations may be in multiple milliseconds, multipath effects may presumably be overcome without much effort.  When digital systems such as EchoLink exist in the communication path, sampling also compromises the accuracy of the channel.  The duration of an element is rounded to conform with sampling.  In the case of $8 kHz$ sampling, the time slots are $125 \mu$s long, and standard deviations in the order of milliseconds should overcome this.  The digital filter used in the receiver software may also have an effect on recovered symbol duration at the receiver.

Our implementation uses a sliding window in the decoder to overcome attenuation and noise in the radio path.  For example, a signal will be thought to exist throughout a window if any point in that window rises above a threshold.  For noisy signals this window size is increased, for example to three carrier frequency cycles in width.  An increased window size decreases the accuracy of the recovered element durations, as the entire window is deduced to be on, or off.  Again, this effect is not major, though when combined with the considerations above a significant deviation may exist.  One may increase element time standard deviation to overcome this, though as previously mentioned this may undermine the security of the channel.  Further, at a certain point the standard deviation may become so high that the use of a PRNG and normal distribution becomes irrelevant, as the derived element times are often either very high or very low with respect to an average or even reasonable value.

While HF, EchoLink, and numbers stations are used here to illustrate the channel, they certainly do not comprise the necessary structure or limitations of our channel.  Alternatives include the use of amateur radio satellites, and reception by web streaming (many repeaters have an online listening feature.)  Numbers stations are a fine illustrative vehicle for this channel, though use for it may certainly be found elsewhere.

\section{Implementation}

Our proof of concept implementation was first written as a general Morse code encoder and decoder, and later extended to include the covert channel described here.  It was written in Python, and is available from http://covert.codes.  Our implementation modifies Morse code symbol time as described above, with some particulars.  When a symbol time is selected from a normal distribution with the pseudo-random duration as its mean, the time is wrapped around via a modulo operation so that it does not fall outside one standard deviation.  A standard deviation above the derived duration conveys a dot, and a standard deviation below conveys a dash.  No change conveys letter demarcation, and durations of two standard deviations from the mean are used to signal the end of a word.

\section{Experiment}
Several local experiments were run at the development site to ensure that the transmitter and receiver were able to communicate covert data over a seemingly innocuous carrier message.  One of these tests is shown after the conclusion.  In this experiment, both dot and dash standard deviations were set to be $10 ms$, dot average length was $60 ms$, and dash average length was $180 ms$.  The trailing noise (characters that occur after the message) may be removed by arranging for a higher, singular high code deviation to indicate {\em end of message}, though the presence of noise after the message deviations have ceased may indicate that the standard deviation was not high enough for the particular test.  In the demonstration test we show the receiver first decoding the carrier Morse code message, then show it also decoding the covert message.  Our software, which can be found at http://covert.codes, is still under development and so replication of this experiment may yield different results.

Comprehensive radio tests were not undertaken due to \cite{part97}, and we request the aid of anyone who can legally test our system over radio paths.  Until radio testing has been undertaken, and the software adjusted in accordance with subsequent findings, this covert channel remains a work in progress.

\section{Conclusion}
We have theorized a covert channel that may be able to convey its covert payload anonymously and quite securely to distant and broad geographical areas.  The channel is flexible, in that it may utilize existing long-range (e.g. HF radio) communication systems, or free systems such as EchoLink to carry its message economically.  Should EchoLink not be an appropriate vehicle for a given implementation, other systems such as IRLP exist that may be able to fill the role.  Other vehicles for the channel (satellite, web streaming) are speculated to be useful as well.  The channel is shown to work in software, where the output of the transmitter is fed directly into the receiver, but meaningful radio tests have yet to be conducted due to \cite{part97}.

\bibliographystyle{plain}
\bibliography{reference}

\clearpage

\begin{lstlisting}
$ ./cwtx.py -o call.wav -m "cq cq cq calling cq this is XXXXXX testing a radio system.  forgive any interruption.  have a good day." -s statfile -c "Mr. Watson come here, I want to see you." -k "secret"
** Saving output to call.wav
** Using coding frequency 900 Hz
** Using sampling frequency 8000 Hz
** Getting statistics from statfile
** Generating audio
** Finished.

$ ./cwstats.py -i call.wav 

** Read 491101 samples from call.wav with sample frequency 8000 Hz and encoding pcm16
** Signal average: 0.000234056981508
** Using dominant frequency: 898 Hz
** Filtering to between 852 and 942 Hz
** Decoding with tolerance: 0.300

Message: cq cq cq calling cq this is XXXXXX testing a radio system. forgive any
interruption. have a good day.

** Finished

$ ./cwstats.py -i call.wav -c statfile -k "secret"
** Read statistics for covert demodulation from statfile

** Read 489533 samples from call.wav with sample frequency 8000 Hz and encoding pcm16
** Signal average: 0.000224008610111
** Using dominant frequency: 898 Hz
** Filtering to between 852 and 942 Hz
** Decoding with tolerance: 0.300

Message: cq cq cq calling cq this is XXXXXX testing a radio system. forgive any
interruption. have a good day.

Decoded: mr. watson come here, i want to see you.                       a   n ee e   em z  my ite n e  nrtw  i k 

** Finished
\end{lstlisting}

\end{document}